\documentstyle[pra,twocolumn,aps]{revtex}

    \input epsf
    \begin{document}
    \draft
    \title{Experimental properties of Bose-Einstein condensates in 1D
optical lattices: Bloch
     oscillations, Landau-Zener tunneling and mean-field effects}
    \author{M.~Cristiani, O.~Morsch, J.H.~M\"uller, D.~Ciampini,
    and E.~Arimondo}
    \address{INFM, Dipartimento di Fisica, Universit\`{a} di Pisa, Via
    Buonarroti 2, I-56127 Pisa, Italy}
    \date{\today}
    \maketitle
    \begin{abstract}We report experimental results on the properties
    of Bose-Einstein condensates in 1D optical lattices. By
    accelerating the lattice, we observed Bloch oscillations of the
    condensate in the lowest band, as well as Landau-Zener (L-Z) tunneling
    into higher bands when the lattice depth was reduced and/or the
    acceleration of the lattice was increased. The dependence of the
L-Z tunneling
    rate on the condensate density was then related to mean-field
effects modifying the effective
    potential acting on the condensate, yielding good agreement with
recent theoretical work.
    We also present several methods for measuring the lattice depth and discuss the effects of the
     micromotion in the TOP-trap
    on our experimental results.

    \end{abstract}

    \narrowtext
    \section{Introduction}
    In a very short time after their first observation,
    Bose-Einstein condensates (BECs) have advanced from being mere
    physical curiosities to the status of well-studied physical
    systems~\cite{inguscio}. A host of diverse and interesting phenomena such as
    collective modes, quantized vortices, and solitons, to name but
    a few, have been extensively investigated and are now well
    understood~\cite{mewes96,jin97,madison00,aboshaeer01,burger99}.
Going from the harmonic potentials used in the
    experiments mentioned above to optical lattices~\cite{lattreview}
constitutes a
    natural extension of the experimental efforts to periodic
    potentials and has opened up new avenues for research. So far,
    experiments using periodic potentials have focused mainly on
    Bragg scattering~\cite{kozuma99,stamperkurn99,ozeri01} and, more
recently, on phase
    properties, involving such intriguing concepts as number
    squeezing~\cite{orzel01} and the Mott insulator
transition~\cite{greiner02,jaksch98}. Some interesting work has
also
    been done on superfluid properties of BECs in optical
    lattices~\cite{burger01,cataliotti01}.

    In the present work we report on experiments with Bose-Einstein
    condensates adiabatically loaded into one-dimensional optical
    lattices~\cite{morsch01,NISTpreprint}. In particular, we look at
the dynamics of the BEC
    when the periodic potential provided by the optical lattice is
    accelerated, leading to Bloch oscillations and L-Z
    tunneling. We then proceed to use L-Z tunneling as a
    tool for measuring the effects of the mean-field interaction
    between the atoms in the condensate. The modification of the
    L-Z tunneling rate in the presence of interactions can
    be interpreted in terms of an effective potential, and we obtain
    good qualitative agreement with a recent theoretical study using this
    approach~\cite{choi99}.

    This paper is organized as follows: After briefly introducing some
essential ideas and terminology
    used in the theoretical treatment of cold atoms in optical lattices
(Sec.~\ref{basics}), we describe
    our experimental apparatus in Sec.~\ref{setup}. After presenting
    in Sec.~\ref{prelim}
    the results of preliminary experiments on the calibration of the
    lattice and on the effects of the condensate micromotion, we turn
to the subject of Bloch oscillations in
    accelerated lattices in Sec.~\ref{Bloch}. The following
    section~\ref{landauzener} deals with L-Z tunneling
    and leads on to a discussion of mean-field effects in
    Sec.~\ref{meanfield_effects}. In Sec.~\ref{conclusions} we
present our
    conclusions and an outlook on further studies, followed by an
    Appendix in which we discuss various parameters relevant to the
    description of our system as an array of tunneling junctions.
    \section{Cold atoms in periodic structures: basic
    concepts}\label{basics}
    The properties of cold atoms in conservative optical lattices (i.e.
far-detuned, so spontaneous scattering is negligible)
     bear a strong resemblance to the behaviour of
    electrons in crystal lattices in condensed
matter physics and
    have, therefore, enjoyed increasing interest since the early
    days of laser cooling. There are a number of excellent review papers on
    the subject~\cite{bendahan96,holthaus00}, so in the following we shall only
    briefly review some basic concepts and establish conventions and
    notations for the remainder of this work.

    The physical system we are considering consists of a Bose-Einstein
condensate in a periodic
    potential created by two interfering linearly polarized laser
    beams with parallel polarizations. The potential seen by the atoms
stems from the ac-Stark
    shift created by an off-resonant interaction between the
    electric field of the laser and the atomic dipole. This results
    in an optical lattice potential of the form
    \begin{equation}
    U(x)=U_0\sin^2(\pi x/d),
    \end{equation}
    where $d$ is the distance between neighbouring wells (lattice constant) and
    \begin{equation}
    U_0=(2/3)\hbar\Gamma(I/I_0)(\Gamma/\Delta),
    \end{equation}
    is the depth of the potential~\cite{bendahan96},
    where $I$ is the intensity of one laser beam, $I_{0}$
is the saturation intensity of the
    $^{87}\mathrm{Rb}$ resonance line, $\Gamma$ is the decay rate of
the first excited state, and $\Delta$ is
    the detuning of the lattice beams from the atomic resonance.
    If the momentum spread of the atoms loaded into such a structure
    is small compared to the characteristic lattice momentum $p_{B}=
2\hbar\pi/d$, then
    their thermal de Broglie wavelength will be large compared to the
    lattice spacing $d$ and will, therefore, extend over many
    lattice sites. The condensates used in our experiment have
    coherence lengths comparable to their spatial extent of
    $\approx 10\,\mathrm{\mu m}$, which should be compared to
    typical lattice spacings in the region of $0.4-1.6\,\mathrm{\mu
    m}$. A description in terms of a coherent delocalized wavepacket within a
    periodic structure is then appropriate and leads us directly to
    the Bloch formalism first developed in condensed matter physics.
As will be explained in the following section,
we could vary the lattice spacing $d$
    by changing the angle between the lattice beams. In this article,
the lattice spacing $d$  always refers to the respective
geometries of the optical lattices used.

    In a lattice configuration in which the two laser beams with
wavevector $k_L$ are counter-propagating, the usual choices of
units
    are the recoil momentum $p_{rec}=\hbar k_L = mv_{rec}$ and the
    recoil energy $E_{rec}=\hbar^2 k_L^2/2m$. In the case of an
    angle-geometry, with the angle $\theta$ between the lattice
    beams (see Fig.~\ref{fig:setup_lat}), it is more intuitive to
    base the natural units on the lattice spacing $d=\frac{\pi}{k_L
    \sin(\theta/2)}$ and the projection $k=\pi/d$ of the laser
wavevector $k_L$ onto the lattice direction. One can then define a
Bloch momentum $p_B =
    2\hbar \pi/d=mv_B$, corresponding to the full extent of the
    first Brillouin zone or, alternatively, to the net momentum
    exchange in the lattice direction between the atoms and the
    two laser beams. Possible choices for the energy unit are either
    the Bloch energy defined as $E_B=\hbar^2 (2\pi)^2/md^2$, or an
    `effective' recoil energy $E_{rec}(\theta)=E_B/8$, where the
    parameter $\theta$ indicates the dependence on the lattice
    geometry~\cite{footnote_zeil}. As an intuitive choice for the
    natural units for the lattice depths is the geometry dependent
    recoil energy $E_{rec}(\theta)$, we shall
    quote the lattice depth $U_0$ in units of this scaled recoil
    energy; for simplicity of notation we shall write $E_{rec}$,
    where it is understood that this always refers to the respective
    lattice geometries. In Sec.~\ref{prelim} on the calibration of the
lattice depth, we also use the parameter
    $s=U_0/E_{rec}$. Throughout the paper, velocities and momenta will be
    quoted in units of the Bloch momentum $p_B$ and Bloch velocity $v_B$.

    In the tight-binding limit ($U_0\gg10\,E_{rec}$), the
    condensate in the lattice can be approximated by wavepackets
    localized at the individual lattice sites (Wannier states). This
description is more intuitive than
    the Bloch picture in the case of experiments in which the
condensate is released from a (deep) optical
    lattice into which it had previously been loaded adiabatically. In
    the present work, this description is only made use of in
    Sec.~\ref{sidepeaks} and in the Appendix, where the Wannier states
are approximated
    by Gaussian functions.

    \section{Experimental Setup}\label{setup}
    Our apparatus for creating Bose-Einstein condensates is
    described in detail in~\cite{jphysbpaper}. Briefly, we use a double-MOT
    setup in order to cool and capture $^{87}\mathrm{Rb}$ atoms and
    transfer them into a time-orbiting potential (TOP) trap.
    Starting with a few times $10^7$ atoms in the magnetic trap, we
    evaporatively cool the atoms down to the critical temperature for
condensation in $\approx
    30\,\mathrm{s}$, obtaining pure condensates containing up to
    $2\times 10^4$ atoms. After condensation, the magnetic trap is
    adiabatically relaxed to mean trap frequencies $\bar{\nu}_{trap}$
on the order of
    $20-60\,\mathrm{Hz}$,  resulting in a
variation of the condensate peak density between
    $2\times 10^{13}\,\mathrm{cm^{-3}}$ and
$10^{14}\,\mathrm{cm^{-3}}$~ .

\begin{figure}
    \centering\begin{center}\mbox{\epsfxsize 2.2 in \epsfbox{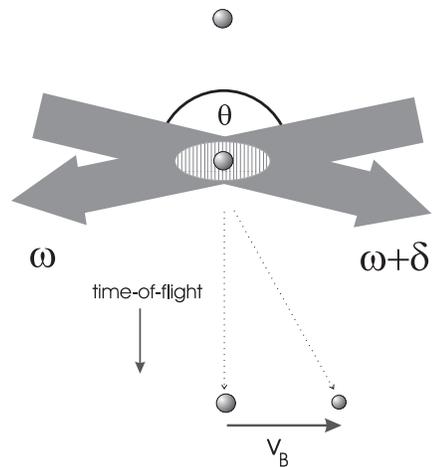}}
    \caption{Schematic of the experimental procedure. The condensate is
loaded into the optical lattice, which
    can transfer momentum to it in units of $p_B=mv_B$. The frequency
difference $\delta$ between the lattice
    beams can be used to create a moving or uniformly accelerated
lattice.}\label{fig:setup_lat}
    \end{center}\end{figure}

    The optical lattice was realized using two linearly polarized
    Gaussian beams (waist $\approx 1.8\,\mathrm{mm}$, maximum power
    $\approx 3\,\mathrm{mW}$) independently controlled by two
    acousto-optic modulators (AOMs) and detuned by about
    $30-60\,\mathrm{GHz}$ above or below the rubidium resonance
    line. The lattice constant $d$ could be varied
    through the angle $\theta$ between the laser beams, as shown in
Fig.~\ref{fig:setup_lat}. Both horizontal and vertical optical
lattices with various
    angles $\theta$ were realized. Furthermore,
    by introducing a frequency difference $\delta$ between the two
    beams, the lattice could be moved at a constant velocity
    $v_{lat}=\frac{\lambda}{2\sin(\theta/2)}\delta$ or accelerated with an
    acceleration
    $a=\frac{\lambda}{2\sin(\theta/2)}\frac{d\delta}{dt}$.
   While in the countere-propagating geometry lattice depths up to $\approx
    2\,E_{rec}$ were realized, in the angle-geometry
lattice depths up to $\approx
    20\,E_{rec}$ could be realized.

    \section{Preliminary Experiments}\label{prelim}
    In order better to understand the effect of the lattice on the
    condensate and to calibrate the theoretically calculated lattice
    depth against experimental values, we performed a series of
    preliminary experiments in conditions in which we expected
    mean-field effects to be negligible, i.e. either with the
    condensate in a weak magnetic trap (after adiabatic relaxation),
    resulting in a low condensate density, or by loading the
    condensate into the lattice after switching off the magnetic
    trap. In the latter case, for horizontal lattice configurations
    the condensate was in free fall after switching off the
    TOP-trap, limiting the interaction time with the lattice to
    $10-15\,\mathrm{ms}$. In all experiments, the condensate was
    observed after $10-20\,\mathrm{ms}$ of time-of-flight by flashing
on a resonant imaging beam for
    $20\,\mathrm{\mu s}$ and observing the shadow cast by the
    condensate onto a CCD-camera.
    \begin{figure}
    \centering\begin{center}\mbox{\epsfxsize 2.8 in \epsfbox{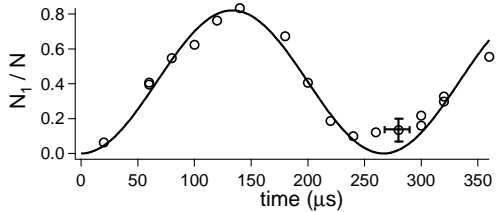}}
    \caption{Rabi oscillations of the condensate in an optical lattice.
Shown here is the fraction $N_1/N$ of condensate
    atoms in the first diffraction order as a function of time. The lattice
     (in the counter-propagating configuration) was moving with
    a constant velocity $v=\frac{1}{2}v_B$. From the Rabi period
$\tau_{Rabi}\approx 260\,\mathrm{\mu s}$
    one finds a lattice depth $U_0\approx 2\,E_{rec}$.}\label{fig:rabi}
    \end{center}\end{figure}
    \subsection{Calibration of the lattice}
    \subsubsection{Rabi
    oscillations}\label{rabiosc}
    If we abruptly switch on an optical lattice moving at a speed
    $\frac{1}{2}v_B$, then in the band-structure picture the condensate
    finds itself at the edge of the Brillouin zone where the first
    and second band intersect at zero lattice depth. Raising the
lattice up to a final depth of $U_0$
    opens up a band-gap of width $\Delta E=U_0/2$ in the shallow lattice
    limit, and hence the populations of the two
bands~\cite{footnote_pops} accumulate a
    phase difference $\Delta \phi = \frac{U_0}{2\hbar} t$, which
    results in the two populations getting back into phase (modulo
    $2\pi$) after a time $\tau_{\rm Rabi} = \frac{2h}{U_0}$.
    In the rest frame of the laboratory, one observes Rabi
    oscillations~\cite{kozuma99} between the momentum classes $|p = 0
\rangle$ and
    $|p =p_B \rangle$ in the shape of varying populations of
    the corresponding diffraction peaks observed after a
    time-of-flight (see Fig.~\ref{fig:rabi}). From the oscillation
    frequency $\Omega_R$ we could then calculate the lattice depth
    $U_0 = 2\hbar \Omega_R$, yielding results that fell short by
    about $20-25$ percent of the calculated value. We attribute the
discrepancy between
    the experimental and theoretical values mainly to the uncertainty in
    our laser intensity measurements and to imperfections in the
    lattice beam cross-section and polarization.\\
    As pointed out in~\cite{bendahan96}, an alternative way of interpreting
    the observed Rabi oscillations in this kind of experiment is to
    consider a two-photon Raman coupling between the two
    momentum states, whose energies differ by $\Delta E = 2\hbar^2
    k^2/m= 4 E_{rec}$. The coupling is resonantly enhanced if the frequency
    difference $\delta$ between the two lattice beams matches $\Delta E/h$,
    i.e. if the lattice velocity $v_{lat}= \frac{\lambda}{2}\delta =
    \frac{\hbar k}{m}= \frac{1}{2}v_B$, as before. The two-photon Rabi
    frequency for the beam intensities corresponding to a lattice depth $U_0$
    can be easily calculated and, again, gives $\Omega_R =
    U_0/2\hbar$.
    \begin{figure}
    \centering\begin{center}\mbox{\epsfxsize 2.8 in \epsfbox{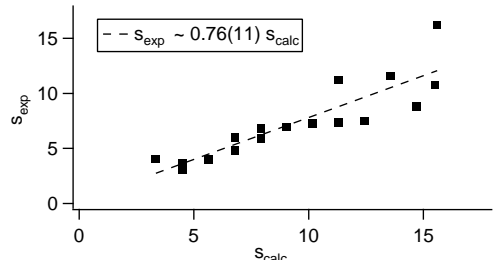}}
    \caption{Calibration of the lattice depth by measuring the
side-peak populations. We
    calculated the lattice depth from the mean value of the plus and
minus first order side-peak using
    Eqn.~\ref{s_func_p1}. By varying the intensity of the lattice
beams, we performed measurements of $s_{exp}$ for
    various values of $s_{calc}$.}\label{sidepeaks}
    \end{center}\end{figure}
    \subsubsection{Analysis of the interference pattern}
    If the depth of a stationary lattice is increased on a timescale
comparable to
    the inverse of the chemical potential of the condensate (in
    frequency units), then the condensate can adiabatically adapt
    to the presence of the periodic potential~\cite{footnote_adiab}.
When the lattice has
    reached its final depth, the system is in a steady state with
    the condensate distributed among the lattice wells (in the limit of
a sufficiently deep lattice in order
     for individual lattice sites to have well-localized wavepackets). If the
    lattice is now switched off suddenly, the individual
    (approximately) Gaussian wavepackets at each lattice site will
    expand freely and interfere with one another. (In this case a
tight-binding approximation
    is more intuitive than a Bloch wave approach.) The resulting
    spatial interference pattern after a time-of-flight of $t$ will be
    a series of regularly spaced peaks with spacing $v_B t$,
corresponding to the various diffraction orders,
     with a
    Gaussian population envelope of width $\approx
\hbar t/m\sigma $, where $\sigma$ is the width of the wavepackets
at the individual lattice sites.
    In particular, Pedri {\em et al.} have shown~\cite{pedri01} that
the relative populations $P_{\pm 1}$ of the
    two symmetric plus and minus first order peaks with respect to
    the zeroth-order central peak are given by $P_{\pm 1} = \exp
    (-4\pi^2 \sigma^2/d^2)$~\cite{footnote_peaks}. For deep
    lattice wells, $\sigma /d$ can be found by making a harmonic
    approximation to the sinusoidal lattice potential about a
    potential minimum, giving a Gaussian width
    $\sigma = d/\pi s^{1/4}$.
    For the relatively shallow lattice used in our experiments
($s<20$), however, this
    approximation is not very accurate. Instead, we used a
    variational ansatz for a Gaussian wavepacket in a sinusoidal
    potential. The resulting transcendental equation can be solved
numerically to
    yield $\sigma$ (see Ref.~\cite{pedri01} and Appendix).
Alternatively, we can find an analytical
     expression for the
    lattice depth as a function of the measured side-peak
    population, giving
    \begin{equation}\label{s_func_p1}
    s_{exp} = \frac{16}{[\ln(P_{\pm 1})]^2}P_{\pm 1}^{-1/4}.
    \end{equation}
    This expression can be used directly to calibrate the lattice
    based on a measurement of the side-peak populations.
    Figure~\ref{sidepeaks} shows the lattice depth $s_{exp}$ as inferred from
    the above equation by measuring the populations of the zeroth
    and plus/minus first order diffraction peaks plotted against the
    lattice depth $s_{calc}$ calculated from the beam intensity and
    detuning, taking into account the losses at the cell windows
    ($\approx 8$ percent). A straight line fit gives $s_{exp} =
    (0.76 \pm 0.1) s_{calc}$, consistent with the results obtained
    by measuring the Rabi oscillation frequency.
    \begin{figure}
    \centering\begin{center}\mbox{\epsfxsize 2.8 in \epsfbox{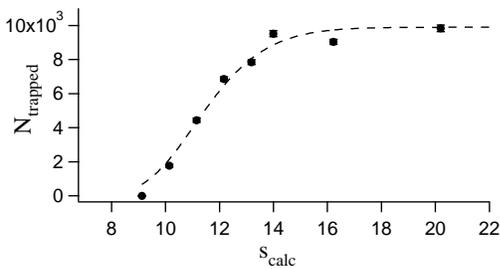}}
    \caption{Calibration of the lattice potential with $d=1.2 \mu$m by
measuring the
tunneling rate. Plotted here as a function of the
    calculated lattice depth $s_{calc}$ is the
    number of condensate atoms $N_{trapped}$ remaining in the lattice at time
    $t_{lat}=10.1\,\mathrm{ms}$ after
    switching off the magnetic trap. Fitting a theoretical curve (see
text) to the data we find
    that $s_{exp}\approx0.76\times s_{calc}$, consistent with the
results of the other calibration techniques.}\label{tunnel_vert}
    \end{center}\end{figure}
    \subsubsection{Tunneling}
    If the lattice beams are arranged such as to create a periodic
    potential along the vertical direction, for deep enough lattice
    potentials the condensate can be held against gravity for
    several hundred milliseconds. As the potential is reduced below
    a critical depth the condensate starts tunneling out of the
    lattice. If this critical depth is fairly small
    ($\lesssim 12\,E_{rec}$), then the tunneling rate can be
    calculated from L-Z theory. If the condensate moving with the
    acceleration $a$ crosses the band gap
    $\Delta E$ at the Brillouin zone edge fast enough, the probability $r$
    for undergoing a transition to the first excited band is
    given by~\cite{zener}
    \begin{equation}
    r=e^{-a_c/a},
    \label{LZ}
\end{equation}
    with the critical acceleration
\begin{equation}
     a_c=\frac{d}{4 \hbar^{2}}(\Delta E)^2.
\label{acceleration}
\end{equation}
If the atoms can tunnel into the second band and, therefore, effectively
    into the continuum, they will no longer be
trapped by the lattice.
    Starting from this assumption, for the condensate accelerated by
    gravity with $a=9.81\,\mathrm{m\,s^{-2}}$ we can
    calculate $N_{trapped}$ the number of atoms that remain trapped after a time
    $t_{lat}$ to be
    \begin{equation}
        N_{trapped}=N_{ini}(1-r)^{t_{lat}/\tau_B},
    \end{equation}
    where $t_{lat}$ is an integer multiple of the Bloch period
$\tau_B=v_B/g$ (see next section). Figure~\ref{tunnel_vert} shows
the
    results of an experiment in which the condensate was loaded into
    a vertical lattice with lattice constant $d=1.2\,\mathrm{\mu m}$,
    after which the magnetic trap was switched off and the number of
    trapped atoms was determined after a time
    $t_{lat}=10.1\,\mathrm{ms}=26\,\tau_B$ as a function of $s_{calc}$. Fitting
    the above equation for $N_{trapped}(t)$ to the data, we found
    that the actual lattice depth was around $75$ percent of the value
    calculated from the lattice beam parameters. This
value agrees with
    those found using Rabi oscillations and side-peak intensity.

     \subsection{Effects of the micromotion}
    The time-orbiting potential trap used in our experiments is an
    intrinsically dynamic trap which relies on the fast rotation of
    the bias field to create an averaged harmonic trapping
    potential. It has been shown, however, that the atoms in
    such a trap perform a small but non-negligible micromotion at
    the rotation frequency $\Omega_{TOP}$ ($2\pi\times
10\,\mathrm{kHz}$ in our set-up) of the bias
field~\cite{prlpaper}. Although
    the spatial amplitude of this fast motion is extremely small
    (less than $100\,\mathrm{nm}$ for typical experimental
    parameters), its velocity can be as large as a few millimeters
    per second. As the Bloch velocities $v_B$ of the lattices used in our
    experiments lie between $3\,\mathrm{mm\,s^{-1}}$ and
    $11.8\,\mathrm{mm\,s^{-1}}$ (depending on the angle $\theta$),
    the micromotion velocity component along the lattice direction
    can be a significant fraction of the Bloch velocity. This has
    two consequences:\\
    (a) If the interaction time of the lattice with the
    condensate is short compared to
    $2\pi/\Omega_{\mathrm{TOP}}=100\,\mathrm{\mu s}$, the initial velocity of
    the condensate is undetermined to within the micromotion
    velocity amplitude.\\
    (b) If, on the other hand, the interaction time with the lattice is long
    compared with $2\pi/\Omega_{\mathrm{TOP}}$, the condensate
    oscillates to and fro in the Brillouin zone and can, if the
    micromotion velocity is large enough, reach the band edge and
    thus be Bragg-reflected. Another possible mechanism is the
parametric excitation at $\Omega_{TOP}$ of transitions to higher
    bands.

    All of these effects are undesirable if one
    wants to conduct well-controlled experiments. In our setup,
    the micromotion takes place in the horizontal plane and was thus
     important when we worked with horizontal
    lattices. In order to minimize the detrimental effects of the
    micromotion, we employed two techniques:\\
    (a) For {\it short interaction times}, we synchronized the instant
at which the lattice
    was switched on with a given phase of
    the rotating bias field. This allowed us to ensure that the lattice was
    always switched on when the condensate velocity along the
    lattice direction was approximately
    zero. A small residual jitter, however,
    was still given by the sloshing (dipole oscillation) of the
    condensate in the magnetic trap, which was especially critical
    when we worked with large lattice constants (and hence small
    Bloch velocities).\\
    (b) For {\it long interaction times}, we phase-modulated one of the
    lattice beams synchronously with the rotating bias field and
    with a modulation depth that resulted in the lattice `shaking'
    with the same velocity amplitude as the micromotion. In this
    way, in the rest-frame of the lattice the condensate was
    stationary (again, save a possible sloshing motion).\begin{figure}
    \centering\begin{center}\mbox{\epsfxsize 2.2 in \epsfbox{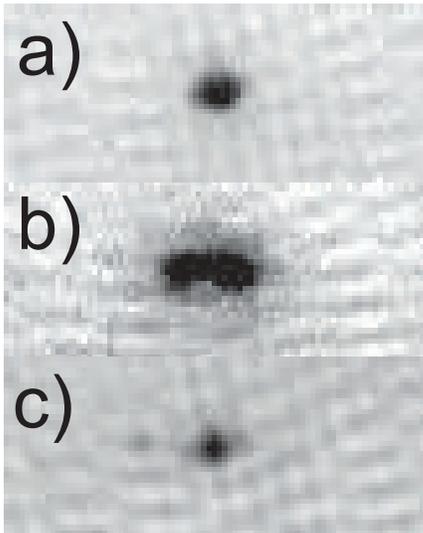}}
    \caption{Effects of the micromotion in a horizontal
counter-propagating lattice configuration. (a) shows a condensate
(after a time-of-flight of
    $15\,\mathrm{ms}$)
    released from
    the magnetic trap with $\overline{\nu}_{trap}=25\,\mathrm{Hz}$ without the
optical lattice. In (b), the optical
    lattice($U_0\approx 2\,E_{rec}$)
    was switched on for $7.5\,\mathrm{ms}$ with the magnetic trap still
on. One clearly sees that the condensate is broadened along the
lattice direction.
    In (c), we compensated the micromotion by phase-modulating one of
the lattice beams (see text). The difference in intensity of the
dark spot in (a) and (c) is due to shot-to-shot fluctuations of
the number of atoms in the condensate. Also, in (c) a faint spot
to the left of the central condensate can be seen; this
corresponds to a part of the condensate having undergone Bragg
reflection due to an initial
sloshing.}\label{fig:micromotion_comp}
    \end{center}\end{figure}

    Figure~\ref{fig:micromotion_comp} shows the effect of method (b) in
    which a condensate was loaded into a lattice ($U_0\approx
    2\,E_{rec}$) with a $1\,\mathrm{ms}$ ramp and then
    left in the lattice for $7.5\,\mathrm{ms}$ before the latter was
suddenly switched off.
    When the micromotion was not compensated by phase-modulating one
    of the lattice beams, the condensate appeared `smeared out' when
    observed after a time-of-flight. Using the compensation
    technique eliminated this effect (Fig.~\ref{fig:micromotion_comp} (c)).

  \begin{figure}
    \centering\begin{center}\mbox{\epsfxsize 2.4 in \epsfbox{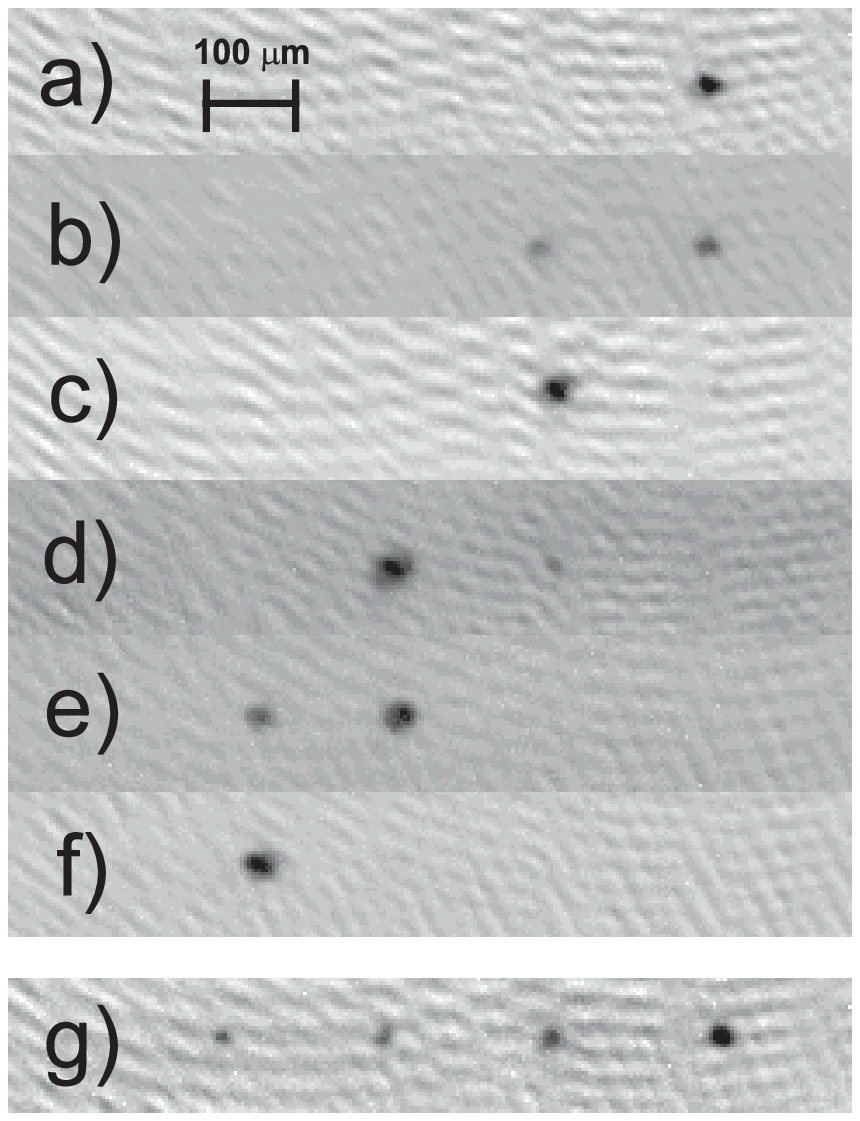}}
    \caption{Acceleration of a Bose-Einstein condensate in the
counter-propagating geometry ($d=0.39\,\mathrm{\mu m}$). In
(a)-(f) the lattice parameters were
    $U_0=2.3\,E_{rec}$ and $a=9.81\,\mathrm{m\,s^{-2}}$, and the
condensate was accelerated for
    $0.1,0.6,1.1,2.1,3.0$ and $3.9\,\mathrm{ms}$, respectively. In (g),
the condensate was accelerated
    for $2.5\,\mathrm{ms}$ with the same lattice depth as above, but with
    $a=25\,\mathrm{m\,s^{-2}}$. In this case, a fraction of the
    condensate underwent L-Z tunneling out of the lowest
    band each time a Bragg-resonance was crossed. Note that the
    separations between the spots vary because detection occurred
    after different times-of-flight.}\label{fig:bloch_labframe}
    \end{center}\end{figure}

    \section{Bloch oscillations}\label{Bloch}
    \subsection{Theoretical considerations}
    One of the most intriguing manifestations of the quantum
    dynamics of particles in a periodic potential are Bloch
    oscillations. Their theoretical explanation is based on the
    evolution in the band-structure picture of a collection of
    particles occupying a small fraction of the Brillouin zone when
    the potential is switched on (meaning that in real space their
    wavefunctions extend over many lattice sites, which translates
     into temperatures well below the recoil temperature
     $T_{rec}=E_{rec}/k_B$, see Sec.~\ref{basics}). If the lattice is
switched on
     adiabatically, then all the atoms will end up in the lowest
     band. Accelerating the atoms by applying a force (real or
     inertial) to them will result in their being moved through the
     Brillouin zone until they reach the band edge. Owing to the
     effects of the periodic potential, at this point there is a gap
     between the first and second band, and unless the acceleration
     is large enough for the atoms to undergo a L-Z
     transition (see Sec.~\ref{landauzener}), they will remain in
     the first band and thus be Bragg-reflected back to the opposite
     end of the Brillouin zone. In the rest-frame of the lattice
     this corresponds to the atoms' velocity oscillating to and fro
     between $+\frac{1}{2}\alpha v_B$ and $-\frac{1}{2}\alpha v_B$
(where $0<\alpha<1$, depending on the
     lattice depth), whereas in the laboratory
     frame these Bloch oscillations manifest themselves as an
     undulating (or, for shallow lattices, almost stepwise) increase
     in velocity rather than the linear increase expected in the
     classical picture in which the atoms are `dragged along' by the
potential. The instantaneous velocity of the atomic wavepacket in
the lattice frame can be calculated from the
     slope of the first band at the corresponding
     quasi-momentum $q$, giving $v= \frac{1}{\hbar} \frac{dE(q)}{dq}$.

     \begin{figure}
    \centering\begin{center}\mbox{\epsfxsize 2.8 in \epsfbox{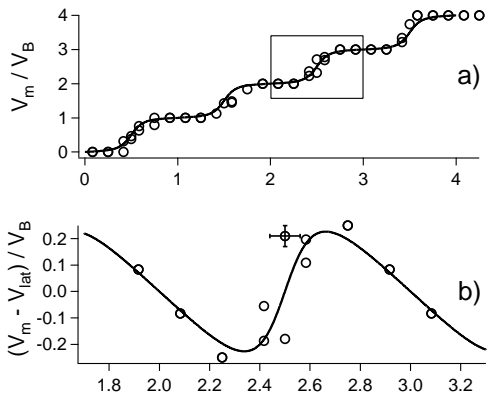}}
    \caption{Bloch oscillations of the condensate mean
    velocity $v_{m}$ in an optical
    lattice. (a) Acceleration
    in the counter-propagating lattice with $d=0.39\,\mathrm{\mu m}$,
$U_0\approx
    2.3\,E_{rec}$ and
    $a=9.81\,\mathrm{m\,s^{-2}}$. Solid line: theory. (b) Bloch oscillations
    in the rest frame of the
    lattice, along with the theoretical prediction (solid line) derived from
    the shape of the lowest
    Bloch band.}\label{fig:bloch_latticeframe}
    \end{center}\end{figure}

    \subsection{Experimental results}
    The condensate was loaded into the (horizontal)
optical lattice with lattice constant $d=0.39\,\mathrm{\mu m}$
    immediately after switching off the magnetic trap. The switch-on
    was done adiabatically with respect to the lattice vibration
    frequencies~\cite{lattreview} $\omega_{vib}
=2E_{rec}\sqrt{s}/\hbar$ (valid for
    $s\gg1$) by ramping up the lattice beam intensity over a time
    $t_{ramp}\approx 100\,\mathrm{\mu s}$. Thereafter, the lattice
    was accelerated with $a=9.81\,\mathrm{m\, s^{-1}}$ by ramping the
    frequency difference $\delta$ between the beams. After a time
    $t_{accel.}$ the lattice was switched off and the condensate was
    observed after an additional time-of-flight of
    $13-18\,\mathrm{ms}$. Figure~\ref{fig:bloch_labframe} shows the results
    of these measurements in the laboratory frame. The Bloch
    oscillations are more evident, however, if one calculates the
    mean velocity $v_m$ as the weighted sum over the momentum components
    after the interaction with the accelerated lattice, as shown in
    Fig.~\ref{fig:bloch_latticeframe}. When the instantaneous lattice
    velocity $v_{lat}$ is subtracted from $v_m$, one clearly sees the
    oscillatory behaviour of $v_m-v_{lat}$. This result is analogous
    to the first observation of Bloch oscillations in cold atoms at
    sub-recoil temperatures~\cite{bendahan96}. The added feature in
    our experiment is that by using a Bose-Einstein condensate released
from a weak magnetic trap,
    the spatial extent of the atomic cloud is sufficiently small so
that after a relatively short time-of-flight
    ($\approx 10-20\,\mathrm{ms}$) the separation between the
individual momentum classes is already
    much larger than the size of the condensate due to its expansion
and can, therefore,
    be easily resolved. The mean
    velocity is then calculated simply by counting the number of atoms in
    each of these classes (the dark dots visible in
    Fig.~\ref{fig:bloch_labframe}).

       \begin{figure}
      \centering\begin{center}\mbox{\epsfxsize 2.8 in
\epsfbox{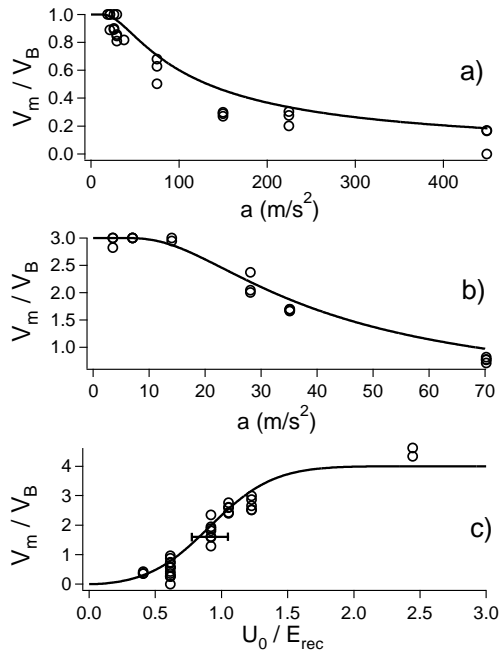}}
    \caption{L-Z tunneling of a condensate in an optical lattice. (a)
and (b): Mean velocity of
    the condensate after acceleration of the lattice to $v_{B}$ and
$3v_{B}$, respectively, as a function of
    acceleration. (c) Mean velocity of the condensate after
acceleration of the lattice to $4.4v_{B}$ as
    a function of lattice depth. In (a) and (b), the lattice depth was
fixed at $U_0=2\,E_{rec}$,
    and in (c) the acceleration $a=9.81\,\mathrm{m\,s^{-1}}$. In (c),
agreement with theory is expected
    to be somewhat less good because the final velocity of the lattice
is not an integer multiple of $v_{B}$ (see
text).}\label{fig:landauzener}
    \end{center}\end{figure}

       \section{L-Z tunneling}\label{landauzener}
    We investigated the $U_{0}$ dependence for the L-Z tunneling of
the condensate
    into the second  band when crossing the
    edge of the Brillouin zone, and  therefore, effectively to the
    continuum, as the gaps between higher bands were negligible for
    the shallow potentials used in our experiments.
    As in Sec.~\ref{Bloch}, we loaded the condensate
    into the optical lattice after switching off the magnetic trap.
    The trap had been adiabatically expanded to a mean trap
    frequency $\overline{\nu}_{trap}\approx 20\,\mathrm{Hz}$ prior to
    switch-off, thus ensuring that the condensate density was small
    ($<10^{13}\,\mathrm{cm^{-3}}$) and, therefore, mean-field
    effects could be neglected. After that, the lattice was
    accelerated to a final velocity $nv_{B}$ ($n=1,2,...$). Each
    time the condensate was accelerated across the
    edge of the Brillouin zone, according to L-Z theory a
    fraction $r$ (see Eqns.~\ref{LZ} and~\ref{acceleration})
underwent tunneling
    into the first excited band. In
    Fig.~\ref{fig:landauzener}, the average velocity of the condensate after the
    acceleration is shown as a function of $a$ and $U_0$ along with
    theoretical predictions using the L-Z tunneling
    probability. If the final lattice velocity is $v_{B}$, one
    finds a final mean velocity $v_m=(1-r)v_{B}$, whereas a
    straightforward generalization of this formula yields
    \begin{equation}
    v_m=v_{B}(1/r-1)[1-(1-r)^n]
    \end{equation}
    for a final lattice velocity of $nv_{B}$. In this case, a fraction
$r$ of the condensate
    undergoes L-Z tunneling each time the Bragg resonance is crossed,
with a remaining fraction
     $1-r$ being accelerated further. Note that this
    result is only independent of the lattice depth (except through
    the tunneling fraction $r$) if the final lattice velocity is
    an integer multiple of $nv_{B}$. As can be seen from
Fig.~\ref{fig:landauzener},
    agreement with theory is good. Again, it should be stressed here
    that owing to the small condensate densities, these measurements
    do not differ qualitatively from those using cold but
    uncondensed atoms~\cite{bendahan96}.

    \begin{figure}
    \centering\begin{center}\mbox{\epsfxsize 2.8 in \epsfbox{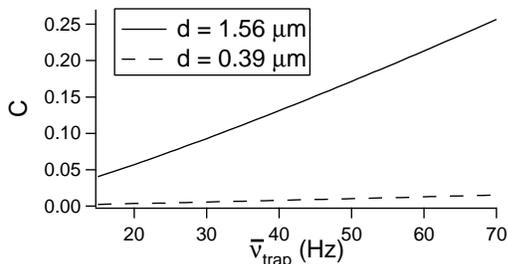}}
    \caption{Calculated dependence of the
    parameter $C$ on
the trap frequency $\bar{\nu}_{trap}$ for
    two different lattice configurations. Mean-field
effects are far more
    important for the larger lattice constant $d=1.56\,\mathrm{\mu m}$,
which results in a $C$ larger by
    a factor of $\approx 16$ with respect to the counter-propagating
configuration ($d=0.39\,\mathrm{\mu m}$). In
    calculating $C$, we assumed a typical condensate number
$N_{tot}=10^4$ and used the Thomas-Fermi
    expression for the condensate peak density.}\label{app_cvsfrq}
    \end{center}\end{figure}
    \section{Mean-field effects}\label{meanfield_effects}
    \subsection{Theoretical considerations}
    In Bose-Einstein condensates, interactions between the
    constituent atoms are responsible for the non-linear behaviour
    of the BEC and can lead to interesting phenomena such as
    solitons~\cite{burger99} and four-wave mixing with matter
    waves~\cite{deng99}. As the atoms are extremely cold, collisions
    between them can be treated by considering only
    $s$-wave scattering, which is described by the scattering length
    $a_s$.  For rubidium the atomic mean-field interaction
    is repulsive corresponding to a positive scattering
    length
    $a_s=5.4\,\mathrm{nm}$.

    For a BEC in an optical lattice, one expects an effect due to
    the mean-field interaction similar to the one responsible for
    determining the shape of a condensate in the Thomas-Fermi limit:
    The interplay between the confining potential and the
    density-dependent mean-field energy leads to a modified ground
    state that reflects the strength of the mean-field interaction.
    Applied to a BEC in a periodic potential, one expects the
    density modulation imposed on the condensate by the potential
    (higher density in potential troughs, lower density where the
    potential energy is high) to be modified in the presence of
    mean-field interactions. In particular, the tendency of the
    periodic potential to create a locally higher density where the
    potential energy of the lattice is low will be counteracted by
    the (repulsive) interaction energy that rises as the local
    density increases.

    The nonlinear interaction of the condensate inside an optical
    lattice with lattice constant $d=\pi/\sin(\theta/2)k_L$ may be
described through a
    dimensionless parameter~\cite{choi99}
    \begin{equation}
    C=\frac{\pi n_0 a_s}{k_L^2sin^2(\theta/2)}=\frac{n_{0}g}{E_{B}},
    \end{equation}
    with $g$ defined in Eqn. (\ref{g}), corresponding to the
    ratio of the nonlinear interaction term and the Bloch energy.
    The parameter $C$ contains the peak condensate density
$n_0$~\cite{footnote_peakdens},
    the scattering length, and the atomic mass $m$.
     In our notation,
the parameter $C$ always
    refers to the respective lattice geometries with angle
    $\theta$. From the dependence of $C$ on the lattice angle $\theta$ it
    follows that a small angle $\theta$ (meaning a large lattice
    constant $d$) will result in a large interaction term $C$. In
    our experiment, creating a lattice with $\theta = 29$ deg (i.e.
    $d=1.56\,\mathrm{\mu m}$)
    allowed us to realize a value of $C$ larger by a factor of more
    than $10$ with respect to Ref.~\cite{anderson98} using a comparable
    condensate density. Figure~\ref{app_cvsfrq} shows our estimates for the
    nonlinear interaction parameter $C$ realized by
    varying the magnetic trap frequency (and, thereby, the
    density $n_{0}$).

    In Ref.~\cite{choi99}, the authors derived an analytical expression
    in the perturbative limit (assuming $U_0\ll E_B$) for the
    effect of the mean-field interaction on the ground state of the
    condensate in the lattice. Starting from the Gross-Pitaevskii equation
    for the condensate wavefunction  in a one-dimensional
    optical lattice ({\it i.e.} a one dimensional Hamiltonian
    equivalent to that of Eq. \ref{hamiltonian}),
    they found that by substituting the
    potential depth $U_0$ with an effective potential
    \begin{equation}
    U_{eff}=\frac{U_0}{1+4C},
    \end{equation}
    the effect of the mean-field interaction could be approximately
    accounted for. This reduction of the effective potential agrees
    with the intuitive picture of the back-action on the periodic
potential of the density
    modulation of the condensate imposed on it by the lattice
    potential. For repulsive interactions, this results in the
    effective potential being lowered with respect to the actual
    optical potential created by the lattice beams.
    \begin{figure}
    \centering\begin{center}\mbox{\epsfxsize 2.8 in \epsfbox{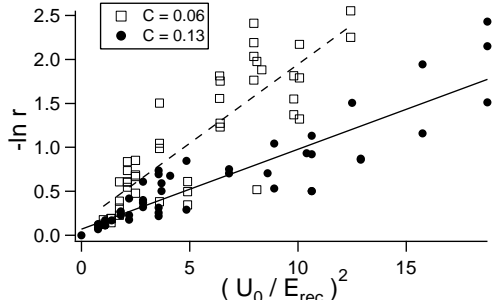}}
    \caption{L-Z tunneling for different values of the interaction
parameter $C$. The tunneling fraction $r$
    was measured by accelerating the condensate to $v_B$ in a vertical
lattice with $d=1.2\,\mathrm{\mu m}$, and $C$ was
    varied by changing the trap frequency and hence the condensate
density $n_0$. The straight lines are best linear fits
    to the data.}\label{fig:meanfield}
    \end{center}\end{figure}
    \subsection{Experimental results}
    In order to measure the effective potential $U_{eff}$, we
    assumed that the perturbative treatment described above
    can be extended to define an effective band gap $\Delta E_{eff}$ at the
    edge of the Brillouin zone which for a particular interaction
    parameter $C$ can be written as
    \begin{equation}\label{effgap}
    \Delta E_{eff} = \frac{U_{eff}}{2}=\frac{U_{0}}{2(1+4C)},
    \end{equation}
    where $\Delta E$ is the
    value of the band gap in the absence of
interactions~\cite{footnote_effgap}. One can
    then derive the interaction parameter $C$ indirectly by
    determining the effective band gap $\Delta E_{eff}$ from the
    L-Z tunneling rate $r$, using Eqns.(\ref{LZ}) and
    (\ref{acceleration})  with the band
    gap $\Delta E_{eff}$.

    Figure~\ref{fig:meanfield} shows the results of measurements of the
    L-Z tunneling rate for two different values of $C$. In
    this experiment, in contrast to those described thus far, the
    lattice was adiabatically ramped up with the magnetic trap
    still on in order to maintain the condensate at a constant density.
     After accelerating the condensate to $v_{B}$, the magnetic trap
and the lattice were both
    switched off and the fraction $r$ that had undergone tunneling (i.e.
the fractional population that
    appeared in the zero momentum class) was measured after a
    time-of-flight. The effective potentials could be derived from the
    slopes of the linear fits in Fig.~\ref{fig:meanfield} and were found to be markedly different.
     Comparing the experimentally determined values for
    $C$ with those calculated on the basis of the experimental
    parameters, we found that the experimental values were larger by
    about a factor of $2.5$. In fact, we expected the predictions
of~\cite{choi99} to be only approximately
    valid (see discussion below).

\begin{figure}
    \centering\begin{center}\mbox{\epsfxsize 2.8 in \epsfbox{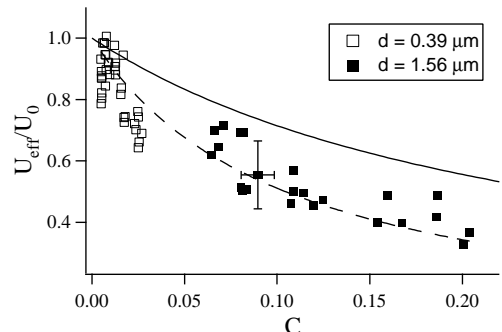}}
    \caption{Dependence of the effective potential $U_{\rm eff}$ on the
    interaction parameter $C$. The different symbols indicate the
     two (horizontal) lattice geometries, with the empty symbols
corresponding to the
     counter-propagating case and the full ones to the angle
     configuration ($d=1.56\,\mathrm{\mu m}$).
     Solid line: theoretical prediction of Choi and Niu's
expression for $U_{eff}$. The parameters in
    these experiments were $a=23.4\,\mathrm{m\,s^{-2}}$ and
    $U_0=2.2\,E_{rec}$ for the counter-propagating lattice and
    $a=3.23\,\mathrm{m\,s^{-2}}$ and $U_0=5.7\,E_{rec}$ for
    the angle geometry.}\label{fig:meanfield_scan}
    \end{center}
\end{figure}

    In order further to test the validity of Eqn.~\ref{effgap}, we
    used two different lattice angles $\theta$ and varied the
    condensate density by changing the trap
frequency\cite{footnote_densscan}. The
effective potential in each case was
     inferred from the tunneling probability $r$ for a fixed lattice
depth. The results
    of these measurements are shown in Fig.~\ref{fig:meanfield_scan}, along
     with the theoretical predictions. Clearly, the reduction of
    $U_{eff}$ with respect to the non-interacting limit is much larger for the
    small lattice angle, as expected from theory. The general
    behaviour of $U_{eff}$ as a function of $C$ is well reproduced
    by our results. Replacing $C$ by $\approx 2.5C$ in the formula
    for $U_{eff}/U_0$ leads to much better agreement with the
    experimental data, which is consistent with the results of
    Fig.~\ref{fig:meanfield}.

    In spite of the good qualitative agreement, we should like to
    point out that the model of Choi and Niu~\cite{choi99} only approximately
    describes our experiment, as it assumes an infinitely extended
    condensate and neglects the radial degrees of freedom of the
    condensate in the one-dimensional lattice. Especially the finite extent of
    the condensate in our experiment (in which only $6-30$ lattice sites
    were occupied by the condensate, see the Appendix) should lead to
non-negligible
    corrections. Also, the analysis of Ref.~\cite{choi99} assumes a
uniform condensate density across
    the entire lattice, whereas in our experiment there was a
pronounced density envelope over the $6-30$ lattice
    wells occupied by the condensate. In the above comparison with
theory we calculated $C$ using the peak condensate
    density.

    \section{Conclusions and outlook}\label{conclusions}
    We have presented experimental results on the adiabatic
    loading and subsequent coherent acceleration of a Bose-Einstein
    condensate in a 1D optical lattice. In the adiabatic
    acceleration limit we have observed Bloch oscillations of the
    condensate mean velocity in the lattice reference frame, whereas
    for larger accelerations and/or smaller lattice depths
    L-Z tunneling out of the lowest band occurred. The
    experimentally observed
    variation of the L-Z tunneling rate with the condensate
    density has been related to the mean-field interaction in the
    condensate leading to a reduced effective potential. Agreement
    with recent theoretical results is satisfactory.\\
    A natural extension of our work on mean-field effects will
    consist in checking theoretical predictions concerning
    instabilities at the edge of the Brillouin zone~\cite{wu01} and
    the possibility of creating bright solitons by exploiting the
    nonlinearity of the Gross-Pitaevskii equation, which can compensate
the negative
    group velocity dispersion at the band edge~\cite{abdullaev01}.

    \section{Acknowledgments}
    This work was
    supported by the INFM through a PRA Project, by MURST through the
COFIN2000 Initiative, and by the
    European Commission through the Cold Quantum-Gases Network, contract
    HPRN-CT-2000-00125. O.M. gratefully acknowledges a
    Marie-Curie Fellowship from the European Commission within the IHP
Programme. The
    authors thank M. Anderlini for help in the data acquisition.

    \appendix
    \section*{An array of coupled wells - relevant parameters}
    The possibility of studying the dynamics of a Bose-Einstein
    condensate spread out coherently over a large number of wells of
    a periodic potential, bearing a close resemblance to an array of
    coupled Josephson junctions, has inspired a host of theoretical
    papers in the past few years. On the experimental side, phase
    fluctuations~\cite{orzel01}, Josephson
    oscillations~\cite{cataliotti01} and the Mott insulator
    transition~\cite{greiner02} have been investigated, invoking
    concepts and notations inherited from the physics of Josephson
    junctions. In order to facilitate the comparison of our work with
    these studies, in this Appendix we report the values pertinent to our
    experiment for the various parameters that are important in the
    description of coherent quantum effects in an array of tunneling
    junctions.

    For the description of a condensate in an array of coupled
    potentials wells, the physical parameters needed to describe the
    dynamics of the system are the on-site interaction $E_C$, and the
    tunneling energy $E_J$. These quantities are defined in a
    variety of ways in the
literature~\cite{javanainen99,giovanazzi00,zapata98}. Our
calculations are based on
     a variational ansatz of the total Hamiltonian
    \begin{equation}
H_{tot}=H_0+g|\Psi(\vec{r})|^2=\frac{-\hbar^2}{2m}\nabla^2+U_0\sin^2(\vec{k_L}\cdot
\vec{r})+g|\Psi(\vec{r})|^2,
    \label{hamiltonian}
\end{equation}
with the interaction parameter $g$ given by
\begin{equation}
     g=4\pi \hbar^2 a_s/m
     \label{g}
     \end{equation}
     and the  wavefunction $\Psi(\vec{r})$ given by
    \begin{equation}
    \Psi(\vec{r})=\sum_n
    \psi_0(x-nd)\sqrt{N_n(t)}e^{i\vartheta_n(t)}\phi(y,z).
    \end{equation}
    Here, $N_n(t)$ is the number of atoms at site $n$,
    $\vartheta_n(t)$ is a site-dependent phase, and $\phi(y,z)$ is the
    part of the wavefunction perpendicular to the lattice direction.
    Basing the variational ansatz for $\psi_0$ on a Gaussian of the form
    $\psi_0(x)=\frac{1}{\sigma^{1/2}\pi^{1/4}}
\exp[-\frac{1}{2}(x/\sigma)^2]$~\cite{footnote_gauss}, we obtain
    a minimum energy wavefunction of width $\sigma$ which, expressed
    in units of the width
$\sigma_{h}=\frac{d}{\pi}(U_0/E_{rec})^{-1/4}$ in the harmonic
approximation
    of the potential wells, satisfies the condition
    \begin{equation}
\exp\left[-\left(\frac{\sigma}{\sigma_h}\right)^2/\sqrt{U_0/E_{rec}}\right]=\left(\frac{\sigma}{\sigma_h}\right)^{-4}.
    \end{equation}
    This equation can be solved numerically to yield $\sigma/\sigma_h$.

    We now define the quantities $E_C$ and $E_J$ as
follows~\cite{footnote_choices}:
    \begin{eqnarray}
    E_C=N_s g_{1\mathrm{D}}\int dx\,\psi_0(x)^4, \\
    E_J=-\int dx\,\psi_0(x)H_0\psi_0(x-d).
    \end{eqnarray}
    In the expression for $E_{C}$, the 1D interaction parameter
    $g_{1\mathrm{D}}$ is defined as
    \begin{equation}
    g_{1\mathrm{D}}=g\frac{1}{\pi \sigma_y \sigma_z},
    \end{equation}
    where
$\sigma_{y,z}$ are the Gaussian widths
     in the $y$ and $z$ directions of
    the radial wavefunction
\begin{equation}
\phi(y,z)=\frac{1}{\pi^{1/2}\sigma_y^{1/2}\sigma_z^{1/2}}\exp[-\frac{1}{2}(y/\sigma_y)^2-\frac{1}{2}(z/\sigma_z)^2].
\end{equation}
    The 1D coupling strength $g_{1\mathrm{D}}$ is equivalent to that
    derived by Olshanii in the case of a cigar shaped atomic
    trap\cite{olshanii}.
$N_s=N_{tot}/n_{occ}$ is the mean number of atoms
    per lattice site, with $N_{tot}=\sum_n N_n$ the total number of
condensate atoms and $n_{occ}$ the number of
    lattice sites occupied, as defined below.

    \begin{figure}
    \centering\begin{center}\mbox{\epsfxsize 2.8 in \epsfbox{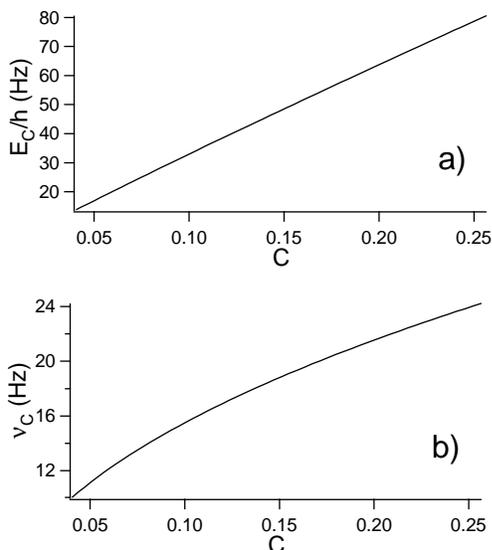}}
    \caption{Dependence of the on-site interaction energy $E_C$,
    in (a), and of the Josephson frequency $\nu_C$, in (b), on the
non-linear parameter $C$ for
    a lattice with $d=1.56\,\mathrm{\mu m}$,
    $U_0=5.6\,E_{rec}$ and $N_{tot}=10^4$. For these parameters, the tunneling
    energy $E_J=h\times8\,\mathrm{Hz}$.}\label{app_ecvsc}
    \end{center}\end{figure}
    \begin{figure}
    \centering\begin{center}\mbox{\epsfxsize 2.8 in \epsfbox{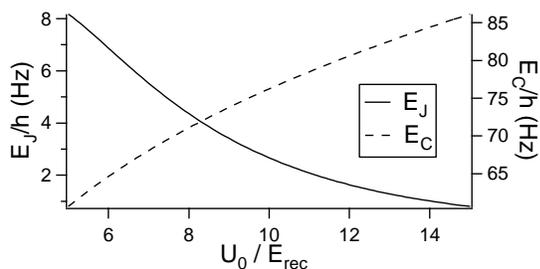}}
    \caption{Variation of $E_J$ and $E_C$ with lattice depth (lattice
constant $d=1.56\,\mathrm{\mu m}$, $N_{tot}=10^4$) for
    a fixed value of the non-linear parameter $C=0.17$.}\label{app_ecejvsuo}
    \end{center}\end{figure}

     \begin{figure}
    \centering\begin{center}\mbox{\epsfxsize 2.8 in \epsfbox{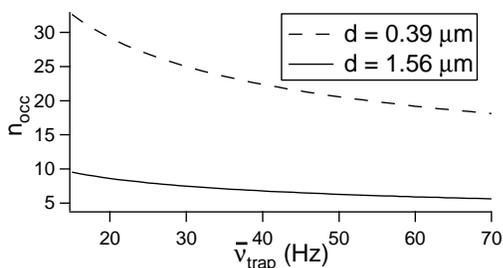}}
    \caption{Number of wells occupied by the condensate as a function
of the magnetic trap
    frequency for two different lattice constants. For the large
lattice constant, the number of wells
    is small, so that finite-size effects are expected to be
important.}\label{app_nowellsvsfrq}
    \end{center}\end{figure}

      As the maximum lattice depth we could experimentally achieve in the
counter-propagating
    configuration was $\approx2\,E_{rec}$, we have only calculated
$E_C$ and $E_J$
    numerically for a lattice in the angle-geometry with
$d=1.56\,\mathrm{\mu m}$, as
    the present model only gives
    reasonable values for $U_0/E_{rec}\gtrsim 4$.
Figure~\ref{app_ecvsc}(a) shows the dependence of the on-site
    interaction energy $E_C$ as a function of the non-linear interaction
    parameter $C$ for a constant lattice depth $U_0=5.6\,E_{rec}$. The
Josephson frequency
    $\nu_c=\sqrt{E_J E_C}/h$ as a function of $C$ is shown in
    Fig.~\ref{app_ecvsc} (b). In Fig.~\ref{app_ecejvsuo}, both $E_J$
and $E_C$ are plotted as
    a function of the lattice depth.

    Finally, we briefly discuss the variation with $\bar{\nu}_{trap}$ of the number
$n_{occ}$ of lattice sites occupied by the condensate. In a rough
approximation, this number is given by the diameter of the
condensate as calculated from the Thomas-Fermi limit divided by
the lattice constant $d$. Pedri {\em  et al.} have used a more
refined model~\cite{pedri01} to derive the expression
    \begin{equation}
n_{occ}=1+\frac{2}{d}\sqrt{\frac{\hbar}{2\pi
m\bar{\nu}_{trap}}}\left(\frac{15}{8\sqrt{\pi}}N_{tot}a_s
    \sqrt{\frac{m\pi\bar{\nu}_{trap}}{\hbar}}\frac{d}{\sigma}\right)^{1/5}.
    \end{equation}

    Figure~\ref{app_nowellsvsfrq} shows $n_{occ}$ as a function of
    $\bar{\nu}_{trap}$ for two different lattice geometries. It is
    clear from this plot that in the angle-geometry, the number of wells
    occupied ($<10$) is small and hence we expect finite-size
    effects to be particularly important in this configuration.


\begin{references}

    \bibitem{inguscio} See reviews by W.~Ketterle
    {\it et al.}, and by E.~Cornell {\it et al.} in {\it Bose-Einstein
    condensation in atomic gases}, edited by M.~Inguscio, S.~Stringari
    and C.~Wieman (IOS Press, Amsterdam) (1999).
    \bibitem{mewes96} M.-O.~Mewes, M.R.~Andrews, N.J.~van Druten,
    D.M.~Kurn, D.S.~Durfee, C.G.~Townsend, and W.~Ketterle, Phys.
    Rev. Lett. {\bf 77}, 988 (1996).
    \bibitem{jin97} D.S.~Jin, M.R.~Matthews, J.R.~Ensher,
    C.E.~Wieman, and E.A.~Cornell, Phys. Rev. Lett. {\bf 78}, 764
    (1997).
    \bibitem{madison00} K.W.~Madison, F.~Chevy, W.~Wohlleben, and
    J.~Dalibard, Phys. Rev. Lett. {\bf 84}, 806 (2000).
    \bibitem{aboshaeer01} J.R.~Abo-Shaeer, C.~Raman, J.M.~Vogels,
    and W.~Ketterle, Science {\bf 292}, 5516 (2001).
    \bibitem{burger99} S.~Burger, K.~Bongs, S.~Dettmer, W.~Ertmer,
    K.~Sengstock, A.~Sanpera, G.V.~Shlyapnikov, and M.~Lewenstein, Phys.
    Rev. Lett. {\bf 83}, 5198 (1999).
    \bibitem{lattreview}For reviews see P.S.~Jessen and I.H.~Deutsch, Adv. At.
    Mol. Opt. Phys.
    {\bf 37}, 95 (1996); G.~Grynberg and C.~Trich\'e, in {\it Coherent and
    Collective
    Interactions of Particles and Radiation Beams} edited by A. Aspect,
    W. Barletta, and R. Bonifacio (IOS Press, Amsterdam) (1996); D.R.~Meacher,
    Cont. Phys. {\bf 39}, 329
    (1998).
    \bibitem{kozuma99} M.~Kozuma, L.~Deng, E.W.~Hagley, J.~Wen,
R.~Lutwak, K.~Helmerson, S.L.~Rolston,
    and W.D.~Phillips, Phys. Rev. Lett. {\bf 82}, 871 (1999);
Yu.B.~Ovchinnikov, J.H.~M\"uller, M.R.~Doery, E.J.D.~Vredenbregt,
K.~Helmerson,
    S.L.~Rolston, and W.D.~Phillips, {\it ibid.} {\bf 83}, 284 (1999);
    J.E.~Simsarian, J.~Denschlag,
    M.~Edwards, C.W.~Clark,
    L.~Deng, E.W.~Hagley, K.~Helmerson, S.L.~Rolston, and
W.D.~Phillips, {\it ibid.}
    {\bf 85}, 2040 (2000).
    \bibitem{stamperkurn99} D.M.~Stamper-Kurn, A.P.~Chikkatur,
    A.~G\"orlitz, S.~Inouye, S.~Gupta, D.E.~Pritchard, and
    W.~Ketterle, Phys. Rev. Lett. {\bf 83}, 2876 (1999).
    \bibitem{ozeri01} R.~Ozeri, J.~Steinhauer, N.~Katz, and
    N.~Davidson, e-print: cond-mat$\backslash$ 0112496 (2001).
    \bibitem{orzel01}  C. Orzel, A.K.~Tuchman, M.L.~Fenselau,
M.~Yasuda, and M.A.~Kasevich, Science {\bf 231}, 2386
    (2001).
    \bibitem{greiner02} M.~Greiner, O.~Mandel, T.~Esslinger,
    T.W.~H\"ansch, and I.~Bloch, Nature {\bf 415}, 6867 (2002).
    \bibitem{jaksch98} D.~Jaksch, C.~Bruder, J.I.~Cirac, C.W.~Gardiner,
and P.~Zoller, Phys. Rev. Lett. {\bf
    81}, 3108 (1998).
    \bibitem{burger01} S.~Burger, F.~Cataliotti, C.~Fort, F.~Minardi,
M.~Inguscio, M.L.~Chiofalo, and
    M.P.~Tosi, Phys. Rev. Lett. {\bf 86}, 4447
    (2001).
    \bibitem{cataliotti01} F.S.~Cataliotti, S.~Burger, C.~Fort,
    P.~Maddaloni, F.~Minardi, A.~Trombettoni, A.~Smerzi, and
    M.~Inguscio, Science {\bf 293}, 843 (2001).
    \bibitem{morsch01} Preliminary results were reported in O.~Morsch,
J.H.~M\"uller, M.~Cristiani,
    D.~Ciampini, and E.~Arimondo, Phys. Rev. Lett. {\bf 87}, 140402
    (2001).
    \bibitem{NISTpreprint} Related experiments have been performed
    by the group of W.D.~Phillips; A. Browaeys, private
    communication (2001).
    \bibitem{choi99} D.~Choi and Q.~Niu, Phys. Rev. Lett. {\bf
    82}, 2022 (1999).
   \bibitem{bendahan96}  M.~Raizen,C.~Salomon, and Q. Niu, Phys. Today {\bf
   50} (7), 30 (1997); E.~Peik, M.~Ben~Dahan, I.~Bouchoule, Y.~Castin, and
C.~Salomon, Phys. Rev.
    A {\bf 55}, 2989 (1997).
    \bibitem{holthaus00} M.~Holthaus, J. Opt. B {\bf 2}, 589 (2000).
    \bibitem{footnote_zeil} In Ref.~\cite{keller99}, the authors
    choose $4E_{rec}=E_B/2$ as the natural unit for comparing
    experiments; the factor of $2$ between their unit and our $E_B$
    stems from the alternative definition of $E_{rec}^{alt}=\hbar^2
    k^2/m$ used in many theoretical papers.
    \bibitem{keller99} C.~Keller, J.~Schmiedmayer, A.~Zeilinger,
    T.~Nonn, S.~D\"urr, and G.~Rempe, Appl. Phys. B {\bf 69}, 303
    (1999).
    \bibitem{jphysbpaper} J.H.~M\"uller, D.~Ciampini, O.~Morsch,
G.~Smirne, M.~Fazzi,
    P.~Verkerk, F.~Fuso, and E.~Arimondo,
    J. Phys. B: Atom. Mol. Opt. Phys. {\bf 33}, 4095
     (2000).
    \bibitem{footnote_pops} Initially, the two bands are equally populated because
     a coherent superposition of
    wavefunctions with equal amplitudes in the first and second band
    (i.e. equal populations) yields a flat density distribution on
    the scale of the lattice constant, corresponding to the initial
    condition of a condensate without density modulation.
    \bibitem{footnote_adiab} The various limits of adiabaticity are
    discussed in detail in the literature; see, e.g., Y.B.~Band and
    M.~Trippenbach, e-print: cond-mat/0201123 (2002). In this paper,
    unless otherwise stated,
    by `adiabatic' we mean `in a time that is long compared with the
    inverse of the harmonic oscillator frequency in the lattice
    wells' (on the order of kHz in our experiment).
    \bibitem{pedri01} P.~Pedri, L.~Pitaevskii, S.~Stringari,
    C.~Fort, S.~Burger, F.S.~Cataliotti, P.~Maddaloni, F.~Minardi,
    and M.~Inguscio, Phys. Rev. Lett. {\bf 87}, 220401 (2001).
    \bibitem{footnote_peaks} For our lattice depths ($\lesssim
    20\,E_{rec}$), higher-order diffraction peaks were
    negligible.
    \bibitem{zener} C. Zener, Proc. R. Soc. London Ser A {\bf 137},
    696 (1932).
    \bibitem{prlpaper} J.H.~M\"uller, O.~Morsch, D.~Ciampini,
M.~Anderlini, R.~Mannella,
    and E.~Arimondo, Phys. Rev. Lett. {\bf 85}, 4454 (2000);
J.H.~M\"uller, O.~Morsch, D.~Ciampini, M.~Anderlini, R.~Mannella,
    and E.~Arimondo, C. R. Acad. Sci. Paris, {\bf t.2(IV)}, 649 (2001).
    \bibitem{deng99} L.~Deng, E.W.~Hagley, J.~Wen, M.~Trippenbach, Y.~Band,
    P.S.~Julienne, J.E.~Simsarian, K.~Helmerson, S.L.~Rolston, and
    W.D.~Phillips, Nature {\bf 398}, 6724 (1999).

    \bibitem{footnote_peakdens} We use the peak density calculated
    in the absence of the optical lattice for our determination of
    $C$. In this way, the (slight) modification of the condensate
    density profile by the lattice does not enter into $n_0$ and,
    hence, $C$ is independent of the lattice depth in the shallow
    lattice limit.
    \bibitem{anderson98} B.P.~Anderson and M.~Kasevich, Science {\bf
    282}, 1686 (1998).
    \bibitem{footnote_effgap} We stress here that defining an
    effective band gap $\Delta E_{eff}$ is only a calculational tool
    used to parametrize the variation in the L-Z tunneling rate with
    $C$. Ref.\cite{wu00}  shows that the effect of the
    mean-field interaction is to deform the Bloch bands, thus also
    affecting the tunneling rate.

    \bibitem{wu00} Biao Wu and Qian Niu, Phys. Rev. A {\bf 61},
    023402 (2000).

    \bibitem{footnote_densscan} For the large lattice constant
    ($d=1.56\,\mathrm{\mu m}$), we found that for condensate
    densities $n_0>5\times 10^{13}\,\mathrm{cm^{-2}}$, the two-peaked
    diffraction pattern became smeared out, so that our
    interpretation in terms of tunneling broke down.

    \bibitem{wu01} Biao Wu and Qian Niu, Phys. Rev. A {\bf 64},
    061603 (2001).
    \bibitem{abdullaev01} F.Kh.~Abduallaev, B.B.~Baizakov,
    S.A.~Darmanyan, V.V.~Konotop, and M.~Salerno, Phys. Rev. A {\bf
    64}, 043606 (2001).
    \bibitem{javanainen99} J.~Javanainen, Phys. Rev. A {\bf 60},
    4902 (1999).
    \bibitem{giovanazzi00} S.~Giovanazzi, A.~Smerzi, and S.~Fantoni,
    Phys. Rev. Lett. {\bf 84}, 4521 (2000).
    \bibitem{zapata98} I.~Zapata, F.~Sols, and A.J.~Leggett, Phys.
    Rev. A {\bf 57}, R28 (1998).
    \bibitem{footnote_choices}In our definitions
    of $E_C$ and $E_J$, we follow closely the notation of Ref.~\cite{javanainen99}. Our choice of
    multiplying the on-site interaction integral in $E_C$ by the
average number of atoms per site is
    motivated by the definition in~\cite{javanainen99} of a `local'
chemical potential $\mu$, which
    in our notation is $\mu=2E_C$.
    \bibitem{footnote_gauss} The Gaussian approximation for $\psi_0$
    is reasonably good for moderate lattice depths ($\approx
    5-15\,E_{rec}$). For deeper lattices, the tunneling energy $E_J$
    calculated with Gaussians deviates exponentially from the exact
    expression $E_J=\frac{8E_{rec}}{\pi}\left(U_0/E_{rec}\right)^{3/4}e^{-2\sqrt{U_0/E_{rec}}}$
    valid in the limit of large $U_0/E_{rec}$.
    (Y.~Castin, PhD thesis, Universit\'e Paris VI (1992)).

    \bibitem{olshanii} M. Olshanii, Phys. Rev. Lett. {\bf 81}, 938
    (1998).
    \end{references}
    \end{document}